\documentstyle[prl,twocolumn,aps,epsf]{revtex}
\begin{document}
\draft

\twocolumn[\hsize\textwidth\columnwidth\hsize\csname
@twocolumnfalse\endcsname

\title{\bf    Interlayer  transport in highly anisotropic
  misfit-layer superconductor (LaSe)$_{1.14}$(NbSe$_2$) }
  \author{P.   Szab\'o,$^{1}$  P. Samuely,$^{1}$ J. Ka\v cmar\v cik,$^{1}$
  A.G.M.Jansen,$^{2}$ A. Briggs,$^{3}$ A. Lafond,$^{4}$ and A. Meerschaut$^{4}$}
\address{$^1$Institute of Experimental Physics, Slovak Academy of
Sciences,         SK-04353         Ko\v{s}ice,         Slovakia.\\
$^2$Grenoble       High       Magnetic      Field      Laboratory,
Max-Planck-Institut     f\"{u}r     Festk\"{o}rperforschung    and
Centre National  de la Recherche Scientifique,  B.P. 166, F-38042
  Grenoble Cedex 9, France.\\
$^3$ Centre de Recherche sur  les Tr\` {e}s
  Basses Temp\'{e}ratures,  CNRS, BP 166,  F-38042 Grenoble,
France.\\
$^4$  Institut des  Mat\'eriaux  Jean  Rouxel, BP  32229, F-44322
  Nantes Cedex 03, France.
 }
\date{\today}

\maketitle

\begin{abstract}
  We  show  that  the {\it interlayer} transport in
  a  two-dimensional    superconductor can reveal  a peak
  in the temperature as well  as the magnetic field dependence of
  the  resistivity  near   the  superconducting  transition.  The
  magnetotransport  experiment   was  performed  on   the  highly
  anisotropic             misfit-layer             superconductor
  (LaSe)$_{1.14}$(NbSe$_2$) with
  critical tempertaure $T_c$ of 1.2  K. The effect is interpreted
within  the tunneling mechanism of  the charge transport across
  the Josephson-coupled layers via two
  parallel channels - the quasiparticles and the Cooper pairs. Similar
  behavior can be  found in the high-$T_c$ cuprates  but there it is
  inevitably interfering with the anomalous normal state. The upper
  critical magnetic field can be obtained from the interlayer tunneling conductance.

\end{abstract}

  ]
  Anomalous  transport  properties  of  the  high-$T_c$  cuprates
  remain  unexplained. In  contrast to  the metallic  temperature
  dependence of  the {\it intralayer} resistivity  $\rho _{ab}$, the
  {\it   interlayer}    resistivity   $\rho   _c$    can   reveal
  a semiconducting  behavior  in   some  cases  \cite{ito}.  The
  anisotropy  ratio can  reach a  value $\rho  _c /  \rho _{ab} =
  10^5$    in     the    most    two-dimensional     system    of
  Bi$_2$Sr$_2$CaCu$_2$O$_8$. The  transport in magnetic  field is
  a puzzle  as  well.  The  superconducting  transition  of $\rho
  _{ab}$  broadens considerably  in a  magnetic field  due to the
  complicated behavior of the superconducting vortex matter \cite{blatter} where
  even   melting   of   a   vortex   lattice   can   be  observed
  \cite{schilling}. The  interlayer resistivity  $\rho _c$ as
  a function    of    magnetic    field    $B$    displays   very
  non-conventional  behavior: starting  from the  superconducting
  state  at $T  < T_c$  with increasing  magnetic field  after an
  onset of the  resistivity, a peak appears followed  by a smooth
  decrease  to a  constant $\rho  _c$ at  higher magnetic  fields
  \cite{ando}.
  When the magnetotransport measurement  is performed above $T_c$
  negative magnetoresistance can be observed \cite{vedeneev}.

  Several  models   have  been  proposed  to   explain  the
  interlayer magnetotransport.  Most of them  treat the peak  in
  $\rho _c(B)$  and the following decrease  of the resistivity at
  higher fields  as a consequence  of the anomalous  normal-state
  properties, mainly due to an  existence of the pseudogap in the
  quasiparticle  spectrum   and/or  superconducting  fluctuations
  \cite{pseudogap}.  But,  Gray   and  Kim  \cite{gray}  proposed
  a model where the peak is due  to an interplay of two different
  conductance  channels present  in the  superconducting state of
  the   sample.   The   model   assumes   a   highly  anisotropic
  superconductor as a stack  of weakly coupled internal Josephson
  junctions and the interlayer transport is accomplished by
  tunneling of  quasiparticles and Cooper pairs.  Below the upper
  critical   magnetic   field   due   to   the   opening  of  the
  superconducting  gap in  the quasiparticle  spectrum, $\rho _c$
  increases  but at  a sufficiently  small field  the Cooper pair
  tunneling channel  is opened and  $\rho _c$ decreases  to zero.
  Recently,  Morozov  et  al.  \cite{morozov}  obtained  evidence
  supporting  this  model  in   their  magnetotransport  data  on
  Bi$_2$Sr$_2$CaCu$_2$O$_8$.  However,  in  cuprates  this purely
  superconducting effect is inevitably  complicated by the
  anomalous normal state properties.

  In  the  present  work  we   address the problem  of  the  {\it
  interlayer}    transport   in    the
  misfit-layer  superconductor  (LaSe)$_{1.14}$(NbSe$_2$),
  a quasi two-dimensional system without any
  non-conventional   behavior    in   the   normal   state.
  We show
  that the interlayer transport in the superconducting state of
  this  layered system involves the tunneling
  of quasiparticles and Cooper pairs.

  (LaSe)$_{1.14}$(NbSe$_2$)  is a  low temperature superconductor
  with $T_c$ around 1.2 K belonging to the family of the lamellar
  chalcogenides  \cite{meerschaut1},  where   two  slabs  MX  and
  TX$_2$ are stacked in a  certain sequence. Due to the different
  symmetry of the MX and TX$_2$ layers a misfit results along one
   intralayer   crystallographic  axis  even   if  along  the
  perpendicular   intralayer  axis  a  perfect  fit  of  both
  structures is  achieved \cite{meerschaut1}. In  the case of
  (LaSe)$_{1.14}$(NbSe$_2$)  every intercalated  LaSe layer  with
  the thickness of about 0.6 nm is
  sandwiched  by  one  2H-NbSe$_2$  layer  with  about  the  same
  thickness \cite{nader}. The sandwich unit is
  stabilized by  the electron transfer from the  LaSe to the NbSe$_2$
  slab resulting in the natural  layered system of the insulating
  LaSe and (super)conducting NbSe$_2$ sheets, where the conduction
  is  accomplished by
   the  Nb 4d$_{z^2}$ orbitals \cite{meerschaut1,berner}.

  The title compound  was obtained by the direct  reaction of the
  three  constituents   La/Nb/Se  in  the   stoichiometric  ratios
  \cite{meerschaut1}.  Single   crystals  used  are   of  typical
  dimensions $1 \times 0.8 \times 0.1$~mm$^3$ with $T_c$ = 1.23 K. For the
  interlayer   as  well  as     intralayer  resistivity
  measurements in the  Montgomery configuration \cite{montgomery}
  four
  electrical contacts were prepared at the corners of the top side
  of  the sample  and another  four contacts  in   symmetrical
  mirror  positions at  the corners   of the  bottom side.
  A standard

  \begin{figure}
\epsfverbosetrue
\vspace{10mm}
\epsfxsize=8.2cm
\epsffile{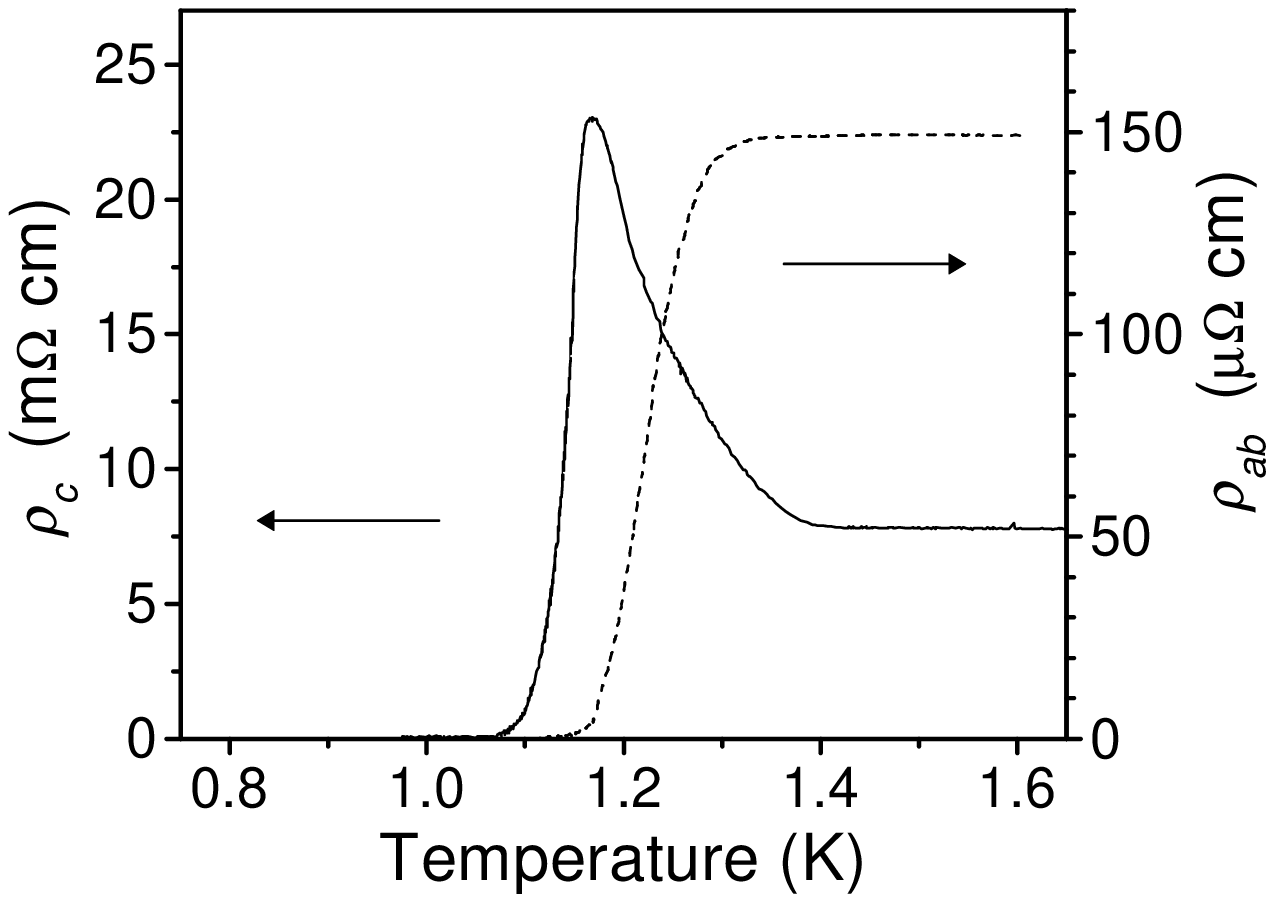}
\vspace{2mm}
\caption{
Temperature   dependence  of   the  {\it   intralayer}
   $\rho  _{ab}$  and  {\it  interlayer}  resistivity
  $\rho _c$ of (LaSe)$_{1.14}$(NbSe$_2$).
}
\end{figure}

\begin{figure}
\epsfverbosetrue
\vspace{10mm}
\epsfxsize=7cm
\epsffile{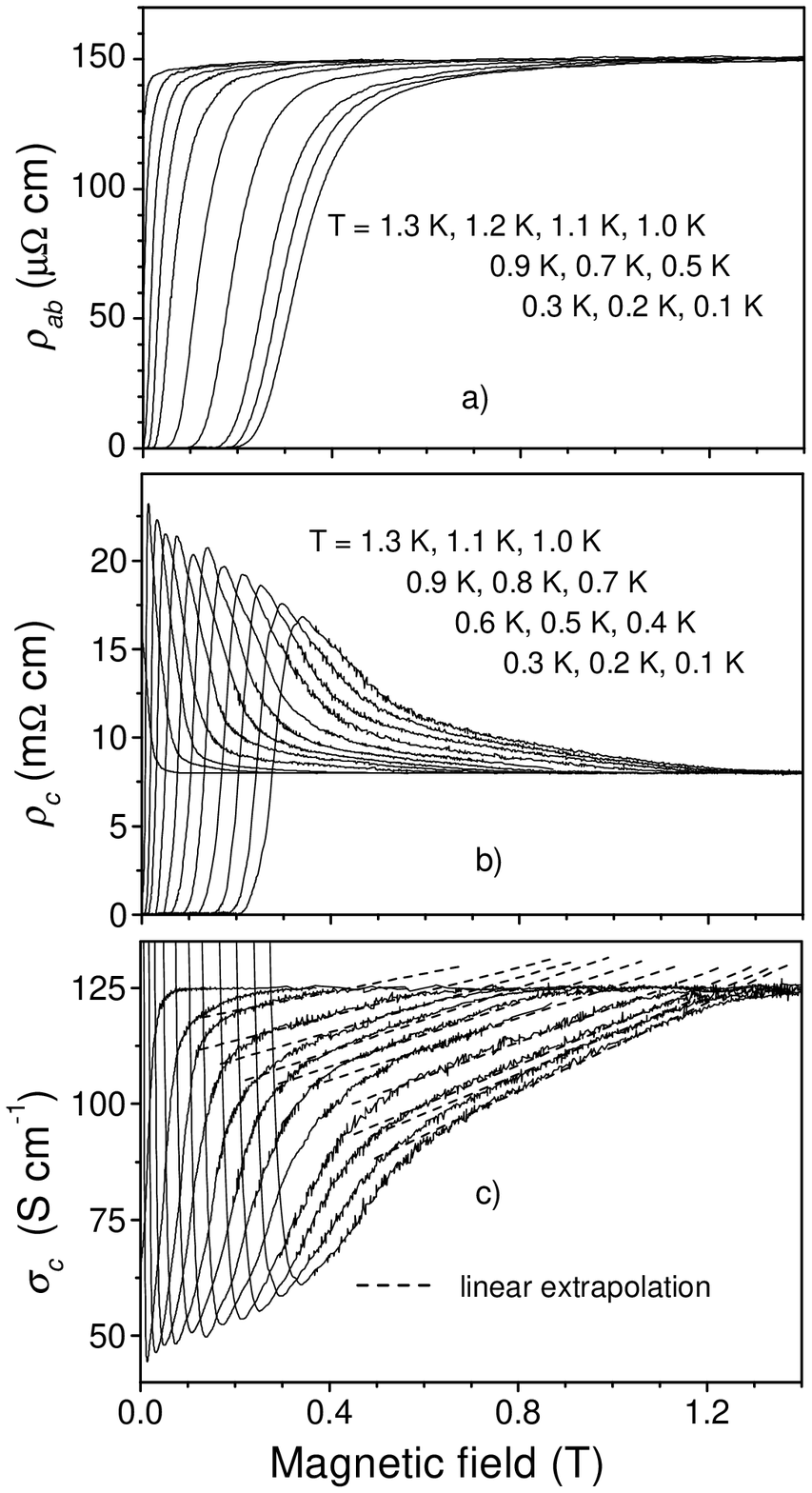}
\vspace{2mm}
\caption{
  a) Intralayer and b)  interlayer magnetoresistive
  superconducting  transitions  at   different  temperatures.
  c) Recalculated conductances from part b).
}
\end{figure}

  \hspace{-4.5mm}   lock-in technique  at 17 Hz was used  to measure the
  temperature  and magnetic  field dependence  of the resistance.
  All measurements  were performed with a  magnetic field applied
  perpendicularly  to  the  planes  -  along  the $c$-axis of the
  samples.  The  field  was   generated  by  a  superconducting
  solenoid  placed  in  the  Ko\v  sice  top-loading refrigerator
  working between 100 mK and 2 K.

  The  temperature dependence  of both $\rho  _c$ and  $\rho  _{ab}$
  revealed  a  metallic  behavior   between  1.5  and  300  K with
  a saturation below  30 K.  The residual resistivity  ratio
  was  about 4 and  the anisotropy ratio  $\rho _c/\rho _{ab}$
  calculated from the  Montgomery configuration about 50
  at 4 K.

  In Fig.~1 the  transition to the  superconducting state is
  shown at zero magnetic field for the  intralayer resistivity
  $\rho  _{ab}$ as  well as  for the  $c$-axis resistivity  $\rho
  _c$.   The     intralayer   resistivity  $\rho_{ab}$  shows
  a conventional transition with a midpoint at $T_c$ = 1.23 K and
  a width  $\Delta  T_c  =  0.1$  K.  This  narrow  single-phase
  transition represents the quality certificate of the sample. On
  the other  hand a very peculiar  transition can be seen  in the
  interlayer  resistivity  $\rho  _c$:  below  1.4  K  the
  resistivity increases by about three  times. Then, below 1.2 K,
  $\rho _c$  drops down reaching the  zero value at about  1.1 K.
  Even with  the lowest current  density (1 mA/cm$^2$)  along the
  $c$-axis  the  zero  resistivity  of  $\rho  _c$  is reached at
  slightly lower temperatures than for $\rho _{ab}$.

  Figure 2 displays the full set of our  intralayer as well as
  interlayer magnetotransport  data measured  at different
  temperatures  from  100  mK  up  to  1.3  K. The  intralayer
  magnetoresistivities (Fig.  2 a) show  conventional transitions
  to the  superconducting state which  are shifted to  higher
  fields  and broadened  as the  temperature is  decreased. Above
  1.4  K the  normal state  is already  achieved in zero magnetic
  field  and no  magnetoresistance  is  observed any  longer. The
  interlayer magnetotransport data are plotted in Fig. 2 b.
  At all  temperatures below 1.4  K the transition  to the normal
  state (with  no magnetoresistance) is  preceded by the  peak.
  Between 1.4 and
  to  1.2~K the resistivity is non zero  at zero
  magnetic field but the peak appears in the field dependence
  with an increasing amplitude.
  Below  1.2~K  the  magnetoresistivity  starts from  zero, the
  following
  peak  is broadened,  its amplitude  decreases and  its position
  shifts  to higher  fields.  The  peak  position in $\rho
  _c(B)$ is  always found in  the range of  magnetic fields where
  the   superconducting   transition   of   the     intralayer
  resistivity  $\rho  _{ab}(B)$  takes  place  at  the respective
  temperature. Figure
  2 c  displays   the   interlayer   magnetotransport  data
  recalculated in the conductance. One can see that in all curves
  below 1 K before reaching  the normal state value  a linear dependence on
  the applied magnetic field is achieved.

  The effect  of the measuring current  density was also examined
  and  the  result  can  be  seen  in  Fig.  3,  where  the
  interlayer resistivity  $\rho _c$ measured at  100 mK is shown
  for  four different  current densities.  The high-field side of the
  peak  in  the  resistivity  is  hardly  affected  unless a high

  \begin{figure}
\epsfverbosetrue
\vspace{3mm}
\epsfxsize=7cm
\hspace{3mm}
\epsffile{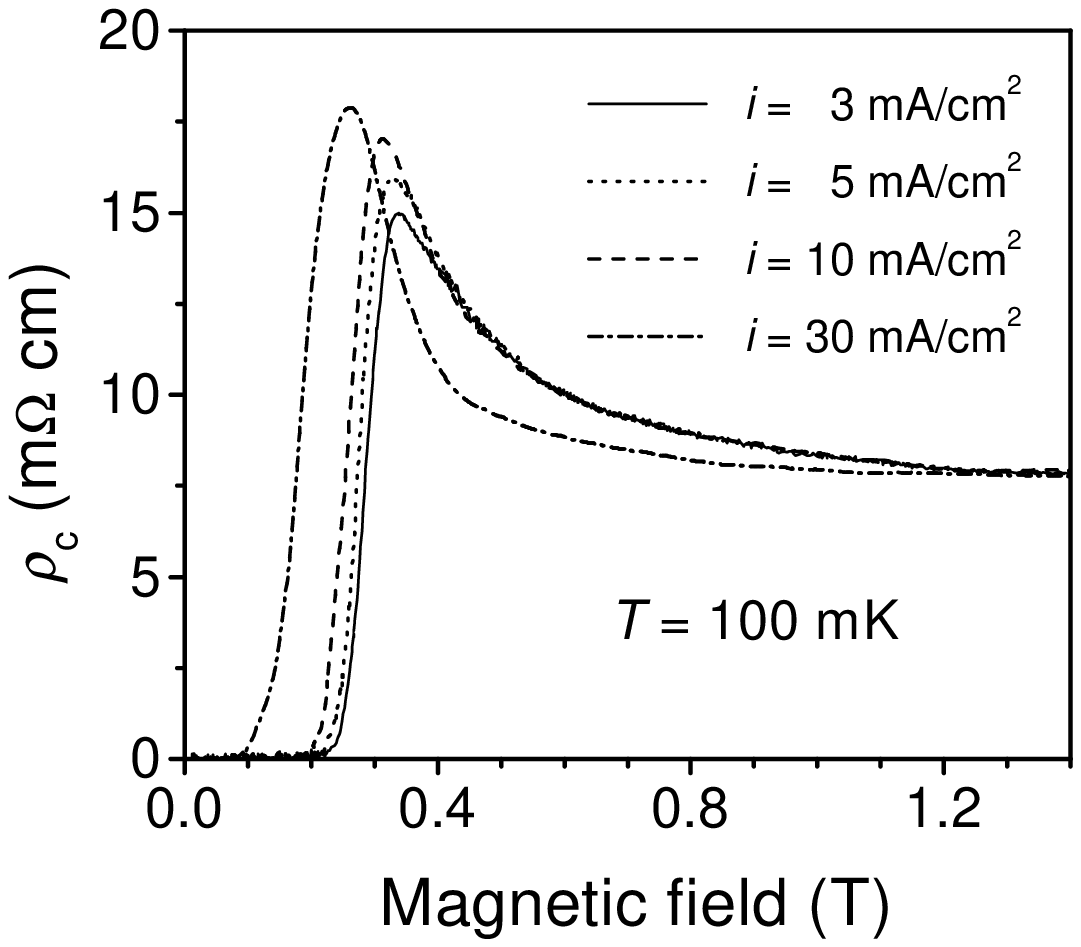}
\vspace{3mm}
\caption{
Effect  of the  measuring current  density on the
  interlayer magnetoresistive superconducting transition.
}
\end{figure}

\begin{figure}
\epsfverbosetrue
\vspace{3mm}
\epsfxsize=7cm
\hspace{3mm}
\epsffile{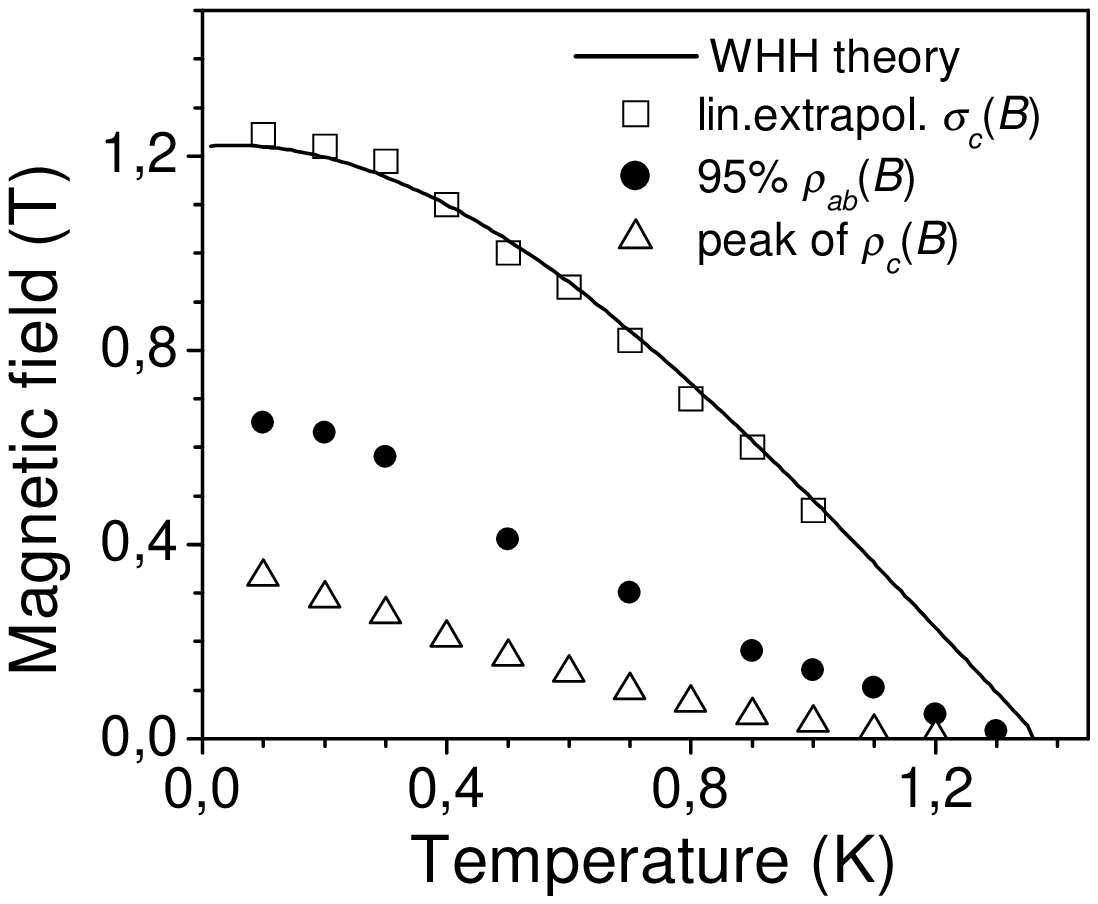}
\vspace{3mm}
\caption{
Critical fields evaluated  from different creteria of the
  superconducting  transition  in  the   intralayer  and
  interlayer resistivities shown in Fig.~2.  The full line shows the
  standard Werthamer-Helfland-Hohenberg temperature dependence
  of the upper critical field $B_{c2}$.
}
\end{figure}

  \hspace{-4.5mm}  current density (30 mA/cm$^2$) is reached where heating affects
  the superconductivity.  But the low-field side of the  peak reveals
  a strong sensitivity  to the current density  - with increasing
  current density  the peak amplitude increases  and its position
  is  shifted to  lower magnetic  fields. Below  3 mA/cm$^2$  the
  current density does not influence the resulting magnetic field
  dependence  of  $\rho  _c$.  The  same  effect of the measuring
  current  density  has  been  observed  in  the  interlayer
  magnetotransport at all temperatures below  $T_c$ as well as in
  the temperature dependence of $\rho _c$ at zero magnetic field.

  The asymmetric effect of the measuring current  density on the peak in the
  interlayer   resistivity   indicates
  to   different  carrier-transport mechanisms
  below and above  the  peak. Moreover  the metallic character  of
  both  intralayer  and interlayer  resistivities
  above  the   superconducting  transition  temperature
  indicates that  the peak effect observed  in the temperature as
  well  as  in  magnetic  field  dependencies  of  the interlayer
  resistivity $\rho  _c$ is related  only to the  superconducting
  transition.  Therefore,  we  will   consider  the  two  channel
  tunneling  model  with  the  quasiparticles  and  Cooper  pairs
  passing   across   the   layers.

  By means of high magnetic field transport measurements for both
  perpendicular  and parallel  field orientation \cite{kacmarcik}
  we have shown that (LaSe)$_{1.14}$(NbSe$_2$) behaves as a quasi
  two-dimensional  system  below  1.1  K.  This  means  that  the
  superconducting coherence length $\xi$ is very anisotropic with
  a value  in  the  $c$-axis  direction  smaller  than  the total
  thickness of the insulating layer and the superconducting layer
  \cite{tinkham}.  We  note  that  the  two-dimensional  behavior
  of  a layered  system is  more pronounced  when the insulating
  layer  thickness  is  bigger  than  the  superconducting  layer
  thickness \cite{deutscher}. As far as the conductance is due to
  the Nb orbitals \cite{berner} the  latter thickness can be very
  small here being restricted just to the atomic thickness of the
  Nb  layer and  the two-dimensional  regime of superconductivity
  would be enhanced.

  Below the  upper critical
  field  $B_{c2}(T)$  a  superconducting  gap  is  opened  in the
  superconducting sheets, which leads to  an increase of the resistivity
  $\rho _c$ at smaller fields as here the quasiparticle tunneling
  channel is  carrying the
  current. At still smaller fields a minimal
  Josephson current  can flow opening a second conductance channel
  which leads to a  rapid decrease of the resistivity. The
  quasiparticle tunneling through internal Josephson junctions is
  obviously  independent  on  the   measuring  current.  But  the
  Josephson  tunneling  of  the  Cooper  pairs  reveals  a strong
  current dependence as is also  observed in our experiment (Fig.
  3). Similarly we can explain  the temperature dependence of the
  interlayer  resistivity $\rho  _c$. Within  this scenario  also
  a slight  shift in  $T_c$'s as  measured by  the intralayer and
  interlayer zero resistivities can be understood: $T_c$ measured
  in  the  plane  represents  the  thermodynamical  value  of the
  material while  $T_c$ measured across the  planes is determined
  by the fact that a measurable Josephson current can flow across
  the whole sample.

  Certain  "critical" magnetic  fields can  be obtained  from the
  magnetic  field dependences  of the  resistivities $\rho _c(B)$
  and $\rho _{ab}(B)$. For  instance, the critical field obtained
  from the  points when the   intralayer resistivity  achieves
  95 per cent of the normal  state value is displayed in the Fig.
  4 by  closed circles.  One can  see a  saturation of the critical
  fields at the lowest temperatures what is typical for the upper
  critical   magnetic   field    $B_{c2}(T)$   of   the   type-II
  superconductors  \cite{werthamer}  but  at  higher temperatures
  there is an anomalous positive  curvature instead of the linear
  temperature dependence.  The latter fact  can be indicative  of
  the presence of vortex melting.  Taking account of the critical
  fluctuations near 100 per cent  of the transition to the normal
  state, we obtain a classical $B_{c2}(T)$ with a linear decrease
  up to  $T_c$ and the zero  temperature extrapolated $B_{c2}(0)$
  equal to 1.2 Tesla \cite{kacmarcik2}.

  A very anomalous  temperature dependence  of the critical
  field  is obtained when
  the  peak in  the  interlayer resistivity-versus-field is
  taken as the upper critical  field position (triangles in
  Fig.~4)    as   is   sometimes    done   in   the    cuprates
  \cite{zavaritsky}.  The  only  criterion  giving  the  expected
  classical $B_{c2}(T)$  dependence is to use  the magnetic field
  where the normal state is reached (squares
 in Fig. 4). These  points can be practically obtained
  from   the  linear   extrapolation  of   the   interlayer
  conductance. It is another  argument supporting the model where
  a peak  in  the    interlayer  resistivity  is  due to the
  interplay  between  the  quasiparticle  and Josephson tunneling
  across  the layers.  In our previous papers
   \cite{samuely} we have shown  that   the  quasiparticle
  tunneling in magnetic fields can give a reliable information on
  the upper  critical fields.  The
  zero-bias tunneling  conductance proportional  to the averaged
  quasiparticle  density  of  states (DOS)  develops  a linear magnetic
  field dependence  near the upper  critical field
  since  the  DOS  is  proportional  to the number of
  vortex cores. In the case of the quasi two-dimensional
  superconductor  (LaSe)$_{1.14}$(NbSe$_2$) the   interlayer
  transport  at higher fields where the Josephson component
  is suppressed is realized via  quasiparticle tunneling and the
  respective  interlayer conductance is  then the zero-bias
  tunneling conductance of the stack of the junctions.

  Finally, in the {\it interlayer} (magneto)transport measured on
  the    quasi     two-dimensional    low-$T_c$    superconductor
  (LaSe)$_{1.14}$(NbSe$_2$) we observed a peak effect in the superconducting
  transition as in the
  cuprates. This phenomenon is observed  regularly in all of many
   samples we measured. The same effect  has also been observed on
  another misfit-layer crystal of (LaSe)$_{1.14}$(NbSe$_2)_2$
  with $T_c \approx 5.7$~K which will be presented elsewhere.
  We have found strong indications that the peak effect is related
  to  the   superconducting  transition  via  the  interplay  of  the
  quasiparticle  and  Cooper-pair  tunneling  mechanisms  of  the
  carrier transport  across the layers.
  The observation of this effect
  in a layered superconductor with conventional normal-state transport
  is of importance
  for the   interpretation of the anisotropic
  transport properties  in the cuprate superconductors.

  \vspace{6cm}

\acknowledgments
  This work has  been supported by the Slovak  VEGA grant No.1148
  and the  liquid nitrogen for  experiment has been  sponsored by
  the U.S. Steel Ko\v sice, DZ Energetika.

\end{document}